\documentclass[12pt]{article}
\usepackage{epsfig}
\usepackage{amsmath}
\usepackage{amssymb}
\usepackage{amscd}
\usepackage{latexsym}
\newcommand{\be}{\begin{equation}}
\newcommand{\ee}{\end{equation}}
\newcommand{\bea}{\begin{eqnarray}}
\newcommand{\eea}{\end{eqnarray}}
\newcommand{\nn}{\nonumber}

\newcommand{\de}{\partial}
\begin{document}
\hfill{\bf CERN-TH/2003-178}\par

\begin{center}
 {\Large\bf\boldmath {Inhomogeneous Color Superconductivity}}
 \rm \vskip1pc {\large
 R. Casalbuoni}\footnote{\uppercase{O}n leave from the \uppercase{D}epartment of
\uppercase{P}hysics of the \uppercase{U}niversity of
\uppercase{F}lorence.}$^,$\footnote{Invited talk at
"QCD@Work 2003 - International Workshop on QCD, Conversano,
Italy, 14--18 June 2003".}\\
\vspace{5mm} {\it{CERN TH-Division, Geneva, Switzerland
\\E-mail: casalbuoni@fi.infn.it}}
 \end{center}


\begin{abstract}We discuss the possibility that in finite
density QCD an anisotropic phase is realized. This case might
arise for quarks with different chemical potential and/or
different masses. In this phase crystalline structures may be
formed. We consider this possibility and we describe, in the
context of an effective lagrangian, the corresponding phonons as
the Nambu-Goldstone bosons associated to the breaking of the space
symmetries.
\end{abstract}



\section{Introduction}

In the last few years  color superconductivity has received a lot
of attention \cite{Rajagopal:2000wf,hsu}. The possibility of
formation of diquark condensates was pointed out since the first
days of QCD \cite{barrois}.   Naively one would expects that, due
to asymptotic freedom, quarks at very high density would form a
Fermi sphere of almost free fermions. However, Bardeen, Cooper and
Schrieffer \cite{BCS} proved that the Fermi surface of free
fermions is unstable in presence of an attractive, arbitrary
small, interaction.  Since in QCD the gluon exchange in the $\bar
3$ channel is attractive one expects the formation of a coherent
state of particle/hole pairs (Cooper pairs). An easy way to
understand the origin of this instability is to remember that, for
free fermions,  the Fermi energy distribution at zero temperature
is given by $f(E)=\theta(\mu-E)$ and therefore the maximum value
of the energy (Fermi energy) is $E_F=\mu$. Then consider the
corresponding grand-potential, $\Omega=E-\mu N$, with $\mu=E_F$.
Adding or subtracting a particle (or adding a hole) to the Fermi
surface does not change $\Omega$, since $\Omega\to (E\pm
E_F)-\mu(N\pm 1)=\Omega$. We see that the Fermi sphere of free
fermions is highly degenerate. This is the origin of the
instability, because if we compare the grand-potential for adding
two free particles or two particles bounded with a binding energy
$E_B$, we find that the difference is given by
$\Omega_B-\Omega=-E_B<0$. Since a bound state at the Fermi surface
can be formed by an arbitrary small attractive interaction
\cite{Cooper}, it is energetically more favorable for fermions to
pair and form condensates.

From the previous considerations it is easy to understand why to
realize superconductivity in ordinary matter is a difficult job.
In fact, one needs an attractive interaction to overcome the
repulsive Coulomb interaction among electrons, as for instance the
one originating from phonon exchange for electrons in metals. On
the other hand the interaction among two quarks in the channel
$\bar 3$ is \underline{attractive}, making color superconductivity
a very robust phenomenon. Notice also that, once taken into
account the condensation effects, in the limit of very high
density one can use asymptotic freedom to get exact results. For
instance, it is possible to get an analytical expression for the
gap \cite{Son:1999uk}. In the asymptotic regime it is also
possible to understand the structure of the condensate
\footnote{At non asymptotic densities other possible condensates
might be formed, see \cite{Alford:2002rz}}. In fact, consider the
matrix element
\begin{equation}
  \langle 0|\psi_{ia}^\alpha\psi_{jb}^\beta|0\rangle
\end{equation}
where $\alpha,\beta=1,2,3$ are color indices, $a,b=1,2$ are spin
indices and  $i,j=1,\cdots, N$ are flavor indices. Its color, spin
and flavor structure is completely fixed by the following
considerations:
\begin{itemize}
\item antisymmetry in color indices $(\alpha,\beta)$ in order to
have attraction; \item antisymmetry in spin indices $(a,b)$ in
order to get a spin zero condensate. The isotropic structure of
the condensate is favored since it allows a better use of the
Fermi surface; \item given the structure in color and spin, Pauli
principles requires antisymmetry in flavor indices.
\end{itemize}
Since the momenta in a Cooper pair are opposite, as the spins of
the quarks (the condensate has spin 0), it follows that the
left(right)-handed quarks can pair only with left(right)-handed
quarks. In the case of 3 flavors the favored condensate is
\begin{equation}
\langle 0|\psi_{iL}^\alpha\psi_{jL}^\beta|0\rangle=-\langle
0|\psi_{iR}^\alpha\psi_{jR}^\beta|0\rangle=
\Delta\sum_{C=1}^3\epsilon^{\alpha\beta C}\epsilon_{ijC}
\end{equation}
This gives rise to the so-called color--flavor--locked (CFL) phase
\cite{Alford:1998mk,Schafer:1998ef}. The reason for the name is
that simultaneous transformations in color and in flavor leave the
condensate invariant. In fact, the symmetry breaking pattern turns
out to be
\[
  SU(3)_c\otimes SU(3)_L\otimes SU(3)_R\otimes U(1)_B\to
  SU(3)_{c+L+R}\nn
\] where $SU(3)_{c+L+R}$ is the diagonal subgroup of the three
$SU(3)$ groups. This is the typical situation when the chemical
potential is much bigger than the quark masses $m_u$, $m_d$ and
$m_s$ (here the masses to be considered are in principle density
depending). However we may ask what happens  decreasing the
chemical potential. At intermediate densities we have no more the
support of asymptotic freedom, but all the model calculations show
that one still has a sizeable color condensation. In particular if
the chemical potential $\mu$ is much less than the strange quark
mass one expects that the strange quark decouples, and the
corresponding condensate should be
\begin{equation}
\langle 0|\psi_{iL}^\alpha\psi_{jL}^\beta|0\rangle=
\Delta\epsilon^{\alpha\beta 3}\epsilon_{ij}
\end{equation}
In fact, due to the antisymmetry in color the condensate must
necessarily choose a direction in color space. Notice that now the
symmetry breaking pattern is completely different from the
three-flavor case. In fact, we have
\[
  SU(3)_c\otimes SU(2)_L\otimes SU(2)_R\otimes U(1)_B\to\] \[\to SU(2)_c\otimes SU(2)_L\otimes
  SU(2)_R\otimes U(1)_B
\]
It is natural to ask what happens in the intermediate region of
$\mu$. It turns out that the interesting case is for $\mu\approx
M_s^2/\Delta$. To understand this point let us consider the case
of two fermions: one massive, $m_1=M_s$ and the other one
massless, $m_2=0$, at the same chemical potential $\mu$. The Fermi
momenta are of course different
\begin{equation}
  p_{F_1}=\sqrt{\mu^2-M_s^2},~~~~p_{F_2}=\mu
\end{equation}
The grand potential for the two unpaired fermions is (the factor 2
comes from the spin degree of freedom) \bea
 \Omega_{\rm unpair.}&=&2\int_{0}^{p_{F_1}}\frac{d^3p}{(2\pi)^3}\left(\sqrt{{\vec p\,}^2+M_s^2}
 -\mu\right)\nn\\&+&
 2\int_{0}^{p_{F_2}}\frac{d^3p}{(2\pi)^3}\left(|\vec p\,|-\mu\right)
\eea In order to pair the two fermions must reach some common
momentum $p_{\rm comm}^F$, and the corresponding grand potential
is
 \bea
 \Omega_{\rm pair.}&=&2\int_{0}^{p_{\rm comm}^F}\frac{d^3p}{(2\pi)^3}
 \left(\sqrt{{\vec p\,}^2+M_s^2}-\mu\right)\nonumber\\&+&
 2\int_{0}^{p_{\rm comm}^F}\frac{d^3p}{(2\pi)^3}\left(|\vec
 p\,|-\mu\right)-\frac{\mu^2\Delta^2}{4\pi^2}
 \label{omega_pair}\eea
where the last term is the energy necessary for the condensation
of a fermion pair \cite{Landau}. The common momentum $p^F_{\rm
comm}$ can be determined by minimizing $\Omega_{\rm pair.}$ with
respect to $p^F_{\rm comm}$, with the result \be p^F_{\rm
comm}=\mu-\frac{M_s^2}{4\mu}\ee It is now easy to evaluate the
difference $\Omega_{\rm unpair.}-\Omega_{\rm pair.}$ at the order
$M_s^4$, with the result \be \Omega_{\rm unpair.}-\Omega_{\rm
pair.}\approx\frac 1{16\pi^2}\left(M_s^4-4\Delta^2\mu^2\right)\ee
We see that in order to have condensation the condition \be
\mu>\frac{M_s^2}{2\Delta}\ee must be realized. The problem of one
massless and one massive flavor has been studied in ref.
\cite{Kundu:2001tt}. However, one can simulate this situation by
taking two massless quarks   with different chemical potentials,
which is equivalent to have two different Fermi spheres. The big
advantage here is that one can use a study made by Larkin and
Ovchinnikov \cite{LOFF1} and Fulde and Ferrel \cite{LOFF2}. These
authors studied the case of a ferromagnetic alloy with
paramagnetic impurities. The impurities produce a magnetic field
which, acting upon the electron spins, gives rise to a different
chemical potential for the two populations of electrons. It turns
out that  it might be energetically favorable to pair fermions
which are close to their respective Fermi surface (LOFF phase).
However, since the Fermi momenta are different, the Cooper pair
cannot have zero momentum and there is a breaking of translational
and rotational invariance. Therefore, a crystalline phase can be
formed. The previous situation is very difficult to be realized
experimentally, but there have been claims of observation of this
phase in heavy-fermion superconductors \cite{gloos} and in
quasi-two dimensional layered organic superconductors \cite{nam}.
The authors of ref. \cite{Alford:2000ze} have extended the
calculation of ref. \cite{LOFF1,LOFF2} to the case of two-flavor
QCD and we will review here their results. Also, since the LOFF
phase can give rise to crystalline structures, phonons are
expected. We will also discuss the effective lagrangians for the
phonons in different crystalline phases and show how to evaluate
the parameters characterizing them, as the velocity of
propagation. Finally we will consider an astrophysical
application.

\section{The LOFF phase}
\begin{figure}[ht]
\centerline{\epsfxsize=2.0in\epsfbox{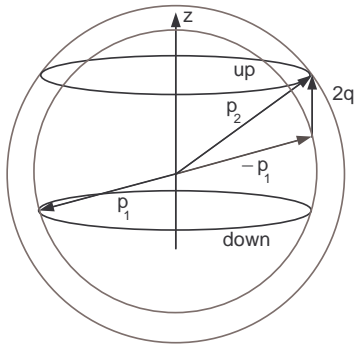}}
\caption{Pairing of fermions belonging to two Fermi spheres of
different radii according to LOFF. \label{fig1}}
\end{figure}

According to the authors of refs. \cite{LOFF1,LOFF2} when fermions
belong to two different Fermi spheres, they  may prefer to pair
staying as much as possible close to their own Fermi surface. When
they are sitting exactly at the surface, the pairing is as shown
in Fig. \ref{fig1}.
We see that the total momentum of the pair is ${\vec p}_1+{\vec
p}_2=2\vec q$ and, as we shall show, $|\vec q\,|$ is fixed
variationally whereas the direction of $\vec q$ is chosen
spontaneously. Since the total momentum of the pair is not zero
the condensate breaks rotational and translational invariance. The
simplest form of the condensate compatible with this breaking is
just a simple plane wave (more complicated functions will be
considered later) \be \langle\psi(x)\psi(x)\rangle\approx\Delta\,
e^{2i\vec q\cdot\vec x}\label{single-wave}\ee It should also be
noticed that the pairs use much less of the Fermi surface than
they do in the BCS case. In fact, in the case considered in Fig.
\ref{fig1} the fermions can pair only if they belong to the
circles in the figure. More generally there is a quite large
region in momentum space (the so called blocking region) which is
excluded from the pairing. This leads to a  condensate smaller
than the BCS one.

Before discussing the LOFF case let us  review the gap equation
for the BCS condensate. We have said that the condensation
phenomenon is the key feature of a degenerate Fermi gas with
attractive interactions. Once one takes into account the
condensation the physics can be described using the Landau's idea
of quasi-particles. In this context quasi-particles are nothing
but fermionic excitations around the Fermi surface described by
the following dispersion relation \be \epsilon(\vec
p,\Delta_{BCS})=\sqrt{\xi^2+\Delta_{BCS}^2}\ee with \be \xi=E(\vec
p)-\mu\approx \frac{\partial E(\vec p)}{\partial \vec
p}\Big|_{\vec p={\vec p}_F}\cdot (\vec p-{\vec p}_F)={\vec
v}_F\cdot(\vec p-{\vec p}_F)\label{xi}\ee The quantities ${\vec
v}_F$ and $(\vec p-{\vec p}_F)$ are called the Fermi velocity and
the residual momentum respectively. A easy way to understand how
the concept of quasi-particles comes about in this context is to
study the gap equation at finite temperature. For simplicity let
us consider the case of a four-fermi interaction. The euclidean
gap equation is given by \be 1=g\int\frac{d^4p}{(2\pi)^4}\frac
1{p_4^2+|\vec p\,|^2+\Delta^2_{BCS}}\ee From this expression it is
easy to get the gap equation at finite temperature. We need only
to convert the integral over $p_4$ into a sum over the Matsubara
frequencies \be 1=gT\int\frac{d^3
p}{(2\pi)^3}\sum_{n=-\infty}^{+\infty}\frac{1}{((2n+1)\pi
T)^2+\epsilon^2(\vec p,\Delta_{BCS})}\ee Performing the sum we get
\be
1=\frac{g}2\int\frac{d^3p}{(2\pi)^3}\frac{1-n_u-n_d}{\epsilon(\vec
p,\Delta_{BCS})}\label{gap_T_finite}\ee Here $n_u$ and $n_d$ are
the finite-temperature distribution functions for the excitations
(quasi-particles) corresponding to the original pairing fermions
\be n_u=n_d=\frac 1{e^{\epsilon(\vec p,\Delta_{BCS})/T}+1}\ee At
zero temperature ($n_u=n_d\to 0$) we find (restricting the
integration to a shell around the Fermi surface)\be
1=\frac{g}2\int\frac{d\Omega_p\, p_F^2\,
d\xi}{(2\pi)^3}\frac{1}{\sqrt{\xi^2+\Delta_{BCS}^2}}\ee In the
limit of weak coupling we get \be \Delta_{BCS}\approx 2\,\bar\xi\,
e^{-2/(g\rho)}\label{BCS}\ee where $\bar\xi$ is a cutoff and \be
\rho=\frac{p_F^2}{\pi^2 v_F}\ee is the density of states at the
Fermi surface. This shows that decreasing the density of the
states the condensate decreases exponentially. From a
phenomenological point of view, one determines the coupling $g$
requiring that the same four-fermi interaction at zero temperature
and density, gives rise to a constituent mass of the order of
$400~MeV$. From this requirement, using  values for $\mu\approx
400\div 500~MeV$ (interesting for the physics of compact stellar
objects),  one obtains values of $\Delta$ in the range $20\div
100~MeV$. Since at very high density it is possible to use
perturbative QCD one can evaluate the gap from first principles
\cite{Son:1998uk}. The result is \be \Delta\approx b\mu g_s^5
e^{-3\pi^2/\sqrt{2}g_s}\ee It is interesting to notice that from
Nambu-Jona Lasinio type of models one would expect a behavior of
the type $\exp(-c/g^2)$ rather than $\exp(-c/g)$. This is due to
an extra infrared singularity from the gluon propagator. Although
this result is strictly valid only at extremely high densities, if
extrapolated down to densities corresponding to $\mu\approx
400\div 500~MeV$, one finds again $\Delta\approx 20\div 100~MeV$.

Let us now consider the LOFF case. For two fermions at different
densities  we have an extra term in the hamiltonian which can be
written as \be H_I=-\delta\mu\sigma_3\label{interaction}\ee where,
in the original LOFF papers \cite{LOFF1,LOFF2} $\delta\mu$ is
proportional to the magnetic field due to the impurities, whereas
in the actual case $\delta\mu=(\mu_1-\mu_2)/2$ and $\sigma_3$ is a
Pauli matrix acting on the two fermion space. According to refs.
\cite{LOFF1,LOFF2} this favors the formation of pairs with momenta
\be \vec p_1=\vec k+\vec q,~~~\vec p_2=-\vec k+\vec q\ee We will
discuss in detail the case of a single plane wave (see eq.
(\ref{single-wave})). The interaction term of eq.
(\ref{interaction}) gives rise to a shift in  $\xi$ (see eq.
(\ref{xi})) due both to the non-zero momentum of the pair and to
the different chemical potential \be \xi=E(\vec p)-\mu\to
E(\pm\vec k+\vec q)-\mu\mp\delta\mu\approx \xi\mp\bar\mu\ee with
\be \bar\mu=\delta\mu-{\vec v}_F\cdot\vec q\ee Here we have
assumed $\delta\mu\ll\mu$ (with $\mu=(\mu_1+\mu_2)/2$) allowing us
to expand $E$ at the first order in $\vec q$ (see Fig.
\ref{fig1}). The gap equation has the same formal expression as in
eq. (\ref{gap_T_finite}) for the BCS case \be
1=\frac{g}2\int\frac{d^3p}{(2\pi)^3}\frac{1-n_u-n_d}{\epsilon(\vec
p,\Delta)}\ee but now $n_u\not=n_d$ \be n_{u,d}=\frac
1{e^{(\epsilon(\vec p,\Delta)\pm\bar\mu)/T}+1}\ee where $\Delta$
is the LOFF gap. In the limit of zero temperature we obtain \be
1=\frac{g}2\int\frac{d^3p}{(2\pi)^3}\frac{1}{\epsilon(\vec
p,\Delta)}\left(1-\theta(-\epsilon-\bar\mu)-\theta(-\epsilon+\bar\mu)\right)\label{gap}\ee
The two step functions can be interpreted saying that at zero
temperature  there is no pairing when $\epsilon(\vec
p,\Delta)<|\bar\mu|$. This inequality defines the so called
blocking region. The effect is to inhibit part of the Fermi
surface to the pairing giving rise a to a smaller condensate with
respect to the BCS case where all the surface is used.

\begin{figure}[ht]
\centerline{\epsfxsize=5.3in\epsfbox{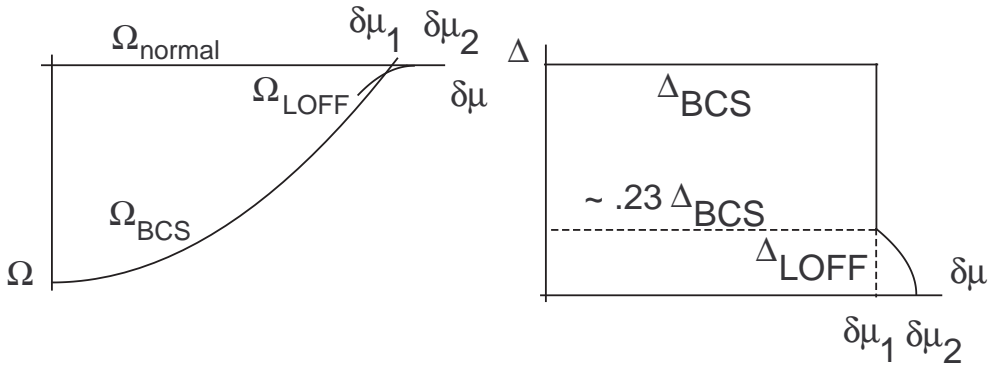}}
\caption{The grand potential and the condensate of the BCS and
LOFF phases vs. $\delta\mu$. \label{fig2}}
\end{figure}

We are now in the position to show that increasing $\delta\mu$
from zero we have first the BCS phase. Then  at
$\delta\mu=\delta\mu_1$ there is a first order transition  to the
LOFF phase \cite{LOFF1,Alford:2000ze},  and at
$\delta\mu=\delta\mu_2>\delta\mu_1$ there is a second order phase
transition to the normal phase (with zero gap)
\cite{LOFF1,Alford:2000ze}. We start comparing the grand potential
in the BCS phase to the one in the normal phase. Their difference
is given by \be \Omega_{\rm BCS}-\Omega_{\rm
normal}=-\frac{p_F^2}{4\pi^2v_F}\left(\Delta^2_{BCS}-2\delta\mu^2\right)\ee
where the first term comes from the energy necessary to the BCS
condensation (compare with eq. (\ref{omega_pair})), whereas the
last term arises from the grand potential of two free fermions
with different chemical potential. We recall also that for
massless fermions $p_F=\mu$ and $v_F=1$. We have again assumed
$\delta\mu\ll\mu$. This implies that there should be a first order
phase transition from the BCS to the normal phase at
$\delta\mu=\Delta_{BCS}/\sqrt{2}$ \cite{clogston}, since the BCS
gap does not depend on $\delta\mu$. The situation is represented
in Fig. \ref{fig2}.
In order to compare with the LOFF phase we will now expand the gap
equation around the point $\Delta=0$ (Ginzburg-Landau expansion)
in order to explore the possibility of a second order phase
transition. Using the gap equation for the BCS phase in the first
term on the right-hand side of eq. (\ref{gap}) and integrating the
other two terms in $\xi$ we get \be \frac
{gp_F^2}{2\pi^2v_F}\log\frac{\Delta_{BCS}}{\Delta}=\frac{gp_F^2}{2\pi^2v_F}
\int\frac{d\Omega}{4\pi}\, {\rm
arcsinh}\frac{C(\theta)}{\Delta}\ee where \be
C(\theta)=\sqrt{(\delta\mu-q v_F\cos\theta)^2-\Delta^2}\ee For
$\Delta\to 0$ we get easily \be
\log\frac{\Delta_{BCS}}{2\delta\mu}=\frac 1 2
\int_{-1}^{+1}\log\left(1-\frac u
z\right),~~~z=\frac{\delta\mu}{qv_F}\label{expansion}\ee This
expression is valid for $\delta\mu$ smaller than the value
$\delta\mu_2$ at which $\Delta=0$. Therefore the right-hand side
must reach a minimum at $\delta\mu=\delta\mu_2$. The minimum is
fixed by the condition \be \frac 1 z\tanh\frac 1 z=1\ee implying
\be qv_F\approx 1.2\, \delta\mu\label{q}\ee Putting this value
back in eq. (\ref{expansion}) we obtain \be \delta\mu_2\approx
0.754\,\Delta_{BCS}\label{deltamu2}\ee From the expansion of the
gap equation around $\delta\mu_2$ it is easy to obtain \be
\Delta^2\approx 1.76\,\delta\mu_2(\delta\mu_2-\delta\mu)\ee
According to ref. \cite{Landau} the difference between the grand
potential in the superconducting state and in the normal state is
given by \be \Omega_{\rm LOFF}-\Omega_{\rm normal}=-\int_0^g\frac
{dg}{g^2}\Delta^2\label{LOFF}\ee Using eq. (\ref{BCS}) and
eq.(\ref{deltamu2}) we can write \be
\frac{dg}{g^2}=\frac{\rho}2\frac{d\Delta_{BCS}}{\Delta_{BCS}}=
\frac{\rho}2\frac{d\delta\mu_2}{\delta\mu_2}\ee Noticing that
$\Delta$ is zero for $\delta\mu_2=\delta\mu$ we are now able to
perform the integral (\ref{LOFF}) obtaining \be \Omega_{\rm
LOFF}-\Omega_{\rm normal}\approx
-0.44\,\rho(\delta\mu-\delta\mu_2)^2\ee We see that in the window
between the intersection of the BCS curve and the LOFF curve in
Fig. \ref{fig2} and $\delta\mu_2$ the LOFF phase is favored.
Furthermore at the intersection there is a first order transition
between the LOFF and the BCS phase. Notice that since
$\delta\mu_2$ is very close to $\delta\mu_1$ the intersection
point is practically given by $\delta\mu_1$. In Fig. \ref{fig2} we
show also the behaviour of the condensates. Altough the window
$(\delta\mu_1,\delta\mu_2)\simeq(0.707,0.754)\Delta_{BCS}$ is
rather narrow, there are indications that considering the
realistic case of QCD \cite{Leibovich:2001xr} the windows may open
up. Also, for different structures than the single plane wave
there is the possibility that the windows opens up
\cite{Leibovich:2001xr}.

\section{Crystalline structures}

\begin{figure}[ht]
\centerline{\epsfxsize=1.9in\epsfbox{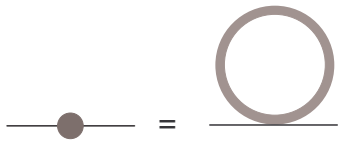}} \caption{Gap
equation in graphical form. The thick line is the exact
propagator. The black dot the gap insertion. \label{fig3}}
\end{figure}

\begin{figure}[ht]
\centerline{\epsfxsize=3.3in\epsfbox{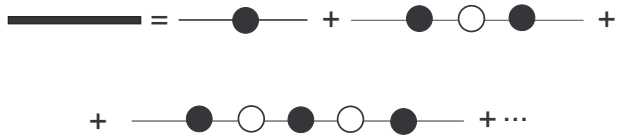}} \caption{The
expansion of the propagator in graphical form. The darker boxes
represent a $\Delta^*$ insertion, the lighter ones a $\Delta$
insertion. \label{fig4}}
\end{figure}

\begin{figure}[ht]
\centerline{\epsfxsize=3.3in\epsfbox{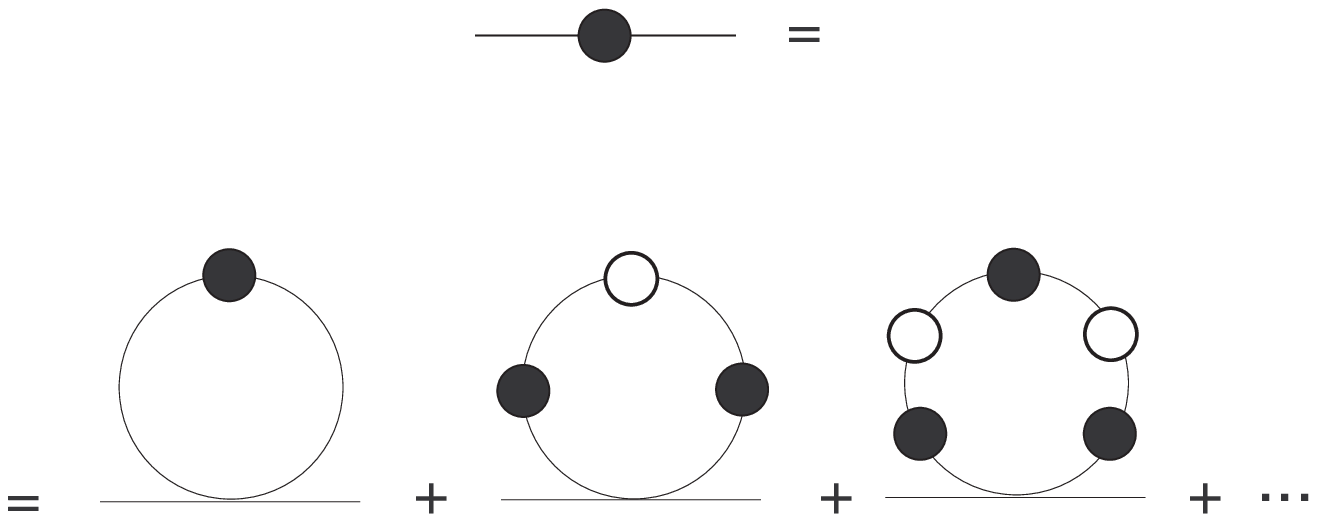}}
\caption{The expansion of the gap equation in graphical form.
Notations as in Fig. \ref{fig4}. \label{fig5}}
\end{figure}

The ground state in the LOFF phase is a superposition of states
with different   occupation numbers ($N$ even)\be
|0\rangle_{LOFF}=\sum_N c_N|N\rangle\ee Therefore the general
structure of the condensate in the LOFF ground state will be \bea
\langle \psi(x)\psi(x)\rangle&=&\sum_N c_N^*c_{N+2}e^{2i\vec
q_N\cdot\vec x}\langle N|\psi(x)\psi(x)|N+2\rangle\nn\\&=& \sum_N
\Delta_N e^{2i\vec q_N\cdot\vec x}\eea The case considered
previously corresponds to all the Cooper pairs having the same
total momentum $2\vec q$. A more general situation, although not
the most general, is when the vectors $\vec q_N$ reduce to a set
$\vec q_i$  defining a regular crystalline structure. The
corresponding coefficients $\Delta_{\vec q_i}$ (linear
combinations of subsets of the $\Delta_N$'s) do not depend on the
vectors $\vec q_i$ since all the vectors belong to the same orbit
of the group. Furthermore all the vectors $\vec q_i$ have the same
lenght \cite{Bowers:2002xr} given by eq. (\ref{q}). In this case
\be\langle 0|\psi(x)\psi(x)|0\rangle=\Delta_q\sum_i e^{2i\vec
q_i\cdot\vec x}\ee This more general case has been considered in
\cite{LOFF1,Bowers:2002xr} by evaluating the grand-potential of
various crystalline structures through a Ginzburg-Landau
expansion, up to sixth order in the gap \cite{Bowers:2002xr}\be
\Omega=\alpha\Delta^2+\frac\beta 2\Delta^4+\frac\gamma
3\Delta^6\label{potential}\ee These coefficients  can be evaluated
microscopically for each given crystalline structure. The
procedure that the authors of ref. \cite{Bowers:2002xr} have
followed is to start from the gap equation represented graphically
in Fig. \ref{fig3}.
Then, they expand the exact propagator in a series of the gap
insertions as given in Fig. \ref{fig4}.
Inserting this expression back into the gap equation one gets the
expansion illustrated in Fig. \ref{fig5}.
On the other hand the gap equation is obtained minimizing the
grand-potential (\ref{potential}), i.e. \be
\alpha\Delta+\beta\Delta^3+\gamma\Delta^5+\cdots=0\ee Comparing
this expression with the result  of Fig. \ref{fig5} one is able to
derive  the coefficients $\alpha$, $\beta$ and $\gamma$.

In ref. \cite{Bowers:2002xr} more than 20 crystalline structures
have been considered, evaluating for each of them  the
coefficients of Eq. (\ref{potential}). The result of this analysis
is that the face-centered cube appears to be the favored structure
among the ones considered (for more details see ref.
\cite{Bowers:2002xr}). However it should be noticed that this
result can be trusted only at $T=0$. In fact one knows that in the
$(\delta\mu,T)$ phase space, the LOFF phase has a tricritical
point and that around this point the favored crystalline phase
corresponds to two antipodal waves (for  reviews see
\cite{Casalbuoni:2003wh,Bowers:2003ye}). Therefore there could be
various phase transitions going down in temperature as it happens
in the two-dimensional case \cite{Shimahara:1998ab}.

\section{Phonons}

QCD at high density is conveniently studied through a hierarchy of
effective field theories. The starting point is the fundamental
QCD lagrangian. The way of obtaining a lower energy effective
lagrangian is to integrate out high-energy degrees of freedom.
Polchinski \cite{Polchinski:1992ed} has shown that  the physics is
particularly simple for energies close to the Fermi energy. In
particular he has shown that all the interactions are irrelevant
except for a four-fermi interaction coupling pair of fermions with
opposite momenta. This is nothing but the interaction giving rise
to the BCS condensation. This physics can be described using the
High Density Effective Theory (HDET) \cite{Hong}. In this theory
the condensation effects are taken into account through the
introduction of a Majorana mass term. The degrees of freedom are
quasi-particles (dressed fermions), holes and gauge fields. This
description is supposed to hold inside a shell around the Fermi
surface of thickness $2\delta$ with  the cutoff $\delta$  smaller
than the Fermi energy but bigger than the condensation energy
$\Delta$, $\Delta\ll\delta\ll E_F$. Going at energies (measured
with respect to the Fermi energy) much smaller  than the gap
energy $\Delta$, all the gapped particles decouple and one is left
with the low energy modes as Goldstone bosons, ungapped fermions
and holes and massless gauged fields according to the breaking
scheme. The corresponding effective theory in the Goldstone sector
can be easily formulated using standard techniques. In the case of
CFL and 2SC such effective lagrangians have been given in refs.
\cite{Casalbuoni:1999wu} and \cite{Casalbuoni:2000cn}. The
parameters of the effective lagrangian can be evaluated at each
step of the hierarchy by matching the Greens functions with the
ones evaluated at the upper level. The same technique will be
applied in this Section to evaluate the propagation parameters of
the Goldstone bosons (phonons) associated to the breaking of
translational and rotational symmetries in the LOFF phase.  The
number and the features of the  phonons depend on the particular
crystalline structure. We will consider here the case of the
single plane-wave \cite{Casalbuoni:2001gt} and of the
face-centered cube \cite{Casalbuoni:2002hr}. We will introduce the
phonons as it is usual for NG bosons \cite{Casalbuoni:2001gt},
that is as the phases of the condensate. Considering the case of a
single plane-wave we introduce a scalar field $\Phi(x)$ through
the replacement \be \Delta(\vec x)=\exp^{2i\vec q\cdot\vec
x}\Delta\to e^{i\Phi(x)}\Delta\label{subst1}\ee We require that
the scalar field $\Phi(x)$  acquires the following expectation
value in the ground state \be \langle\Phi(x)\rangle=2\,\vec
q\cdot\vec x\label{expectation}\ee The phonon field is defined as
\be\frac 1 f\phi(x)=\Phi(x)-2\vec q\cdot\vec x\label{phonon}\ee
Notice that the phonon field transforms nontrivially under
rotations and translations. From this it follows that non
derivative terms for $\phi(x)$ are not allowed. One starts with
the most general invariant lagrangian for the field $\Phi(x)$ in
the low-energy limit. This cuts the expansion of $\Phi$ to the
second order in the time derivative. However one may have an
arbitrary number of space derivative, since from eq.
(\ref{expectation}) it follows that the space derivatives do not
need to be small. Therefore \be L_{\rm phonon}
=\frac{f^2}2\left({\dot\Phi}^2+\sum_k
c_k\Phi(\vec\nabla^2)^k\Phi\right)\ee Using the definition
(\ref{phonon}) and keeping the space derivative up to the second
order (we can make this assumption for the phonon field) we find
\be {L_{\rm phonon}}=\frac 1
2\left({\dot\phi}^2-v_\perp^2\vec\nabla_\perp\phi\cdot\vec\nabla_\perp\phi-
v_\parallel^2\vec\nabla_\parallel\phi\cdot\vec\nabla_\parallel\phi\right)\ee
where \be\vec\nabla_\parallel=\vec n(\vec n\cdot\vec\nabla),~~~~
\vec\nabla_\perp=\vec\nabla-\vec\nabla_\parallel,~~~~\vec
n=\frac{\vec q}{|\vec q\,|}\ee We see that the propagation of the
phonon in the crystalline medium is anisotropic.

\begin{figure}[ht]
\centerline{\epsfxsize=2.5in\epsfbox{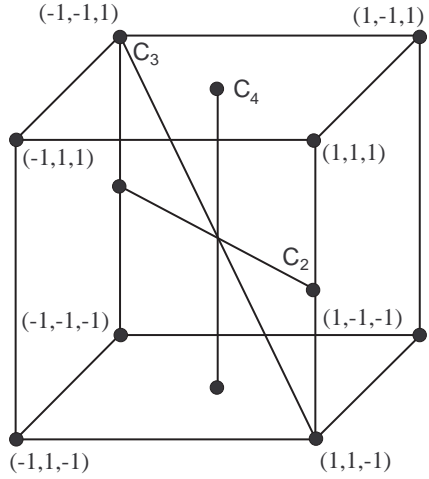}} \caption{The
figure shows the vertices and corresponding coordinates of the
cube described in the text. Also shown are the symmetry axis.
\label{fig6}}
\end{figure}

The same kind of considerations can be made in the case of the
cube. The cube is defined by 8 vectors $\vec q_i$ pointing from
the origin of the coordinates to the vertices of the cube
parameterized as in Fig. \ref{fig6}.

The condensate is given by \cite{Bowers:2002xr} \be
\Delta(x)=\Delta\sum_{k=1}^8e^{2i\vec q_k\cdot\vec
x}=\Delta\sum_{i=1,(\epsilon_i=\pm)}^3e^{2i|\vec q\,|\epsilon_i
x_i}\ee We introduce now three scalar fields such that
\be\langle\Phi^{(i)}(x)\rangle=2|\vec q\,|x_i\ee through the
substitution
\be\Delta(x)\to\Delta\sum_{i=1,(\epsilon_i=\pm)}^3e^{i\epsilon_i\Phi^{(i)}(x)}
\label{phonon_cube}\ee and the phonon fields \be \frac 1 f
\phi^{(i)}(x)=\Phi^{(i)}(x)-2|\vec q\,|x_i\ee Notice that the
expression (\ref{phonon_cube}) is invariant under the symmetry
group of the cube acting upon the scalar fields $\Phi^{(i)}(x)$.
This group has three invariants for the vector representation \bea
&I_2(\vec X)=|\vec X\,|^2, ~I_4(\vec X)=X_1^2 X_2^2 +X_2^2
X_3^3+X_3^2 X_1^2,&\nn\\&I_6(\vec X)=X_1^2 X_2^2 X_3^2&\eea
Therefore the most general invariant lagrangian  under rotations,
translations and the symmetry group of the cube, at the lowest
order in the time derivative, is \bea  L_{\rm phonon}&=&\frac
{f^2} 2\sum_{i=1,2,3}({\dot\Phi}^{(i)})^2\nn\\&+& L_{\rm
s}(I_2(\vec\nabla\Phi^{(i)}),
I_4(\vec\nabla\Phi^{(i)}),I_6(\vec\nabla\Phi^{(i)}))\eea Expanding
this expression at the lowest order in the space derivatives of
the phonon fields one finds \cite{Casalbuoni:2002hr} \bea L_{\rm
phonos} &=&\frac 1 2\sum_{i=1,2,3}({\dot\phi}^{(i)})^2-\frac a 2
\sum_{i=1,2,3}|\vec\nabla\phi^{(i)}|^2\nn\\&-& \frac b 2
\sum_{i=1,2,3}(\de_i\phi^{(i)})^2-
c\sum_{i<j=1,2,3}\de_i\phi^{(i)}\de_j\phi^{(j)}\eea

The parameters appearing in the phonon lagrangian can be evaluated
following the strategy outlined in
\cite{Casalbuoni:2002pa,Casalbuoni:2002my}. One starts from the
QCD lagrangian and derives an effective lagrangian describing
fermions close to the Fermi surface, that is at momenta such that
$p\approx p_F$ but $p-p_F\gg\Delta$. The relevant degrees of
freedom are the fermions dressed by the interaction, the so-called
quasi-particles \cite{Hong}. Going closer to the Fermi surface the
gapped quasi-particles decouple and one is left with the light
modes as NG bosons, phonons and un-gapped fermions. It is possible
to derive the parameters of the last description by the one in
terms of quasi-particles evaluating the self-energy of the phonons
(or the NG bosons) through one-loop diagrams due to fermion pairs.
The couplings of the phonons to the fermions are obtained noticing
that the gap acts as a Majorana mass for the quasi-particles.
Therefore the couplings originate from the substitutions
(\ref{subst1}) and (\ref{phonon_cube}). In this way one finds the
following results: for the single plane-wave \be v_\perp^2=\frac 1
2\left(1-\left(\frac{\delta\mu}{|\vec q\,|}\right)^2\right),~~~
v_\parallel^2=\left(\frac{\delta\mu}{|\vec q\,|}\right)^2\ee and
for the cube \be a=\frac 1 {12},~~~b=0,~~~c=\frac
1{12}\left(3\left(\frac{\delta\mu}{|\vec
q\,|}\right)^2-1\right)\ee

\section{Astrophysical consequences}

A typical phenomenon of the pulsars are the glitches (for a review
see \cite{Carter:1998ji}), that is sudden jumps in the period of
the star. If pulsars are neutron stars with a dense metallic
crust, the effect is explained assuming that some angular momentum
is stored in the vortices  present in the inner neutron
superfluid. When the period of the star slows down , the vortices,
which are pinned to the crystalline crust, do not participate in
the slowing-down until they become unstable releasing suddenly the
angular momentum. Since the density in the inner of a star is a
function of the radius, it results that one has a sort of
laboratory to study the phase diagram of QCD at zero temperature,
at least in the corresponding range of densities. A possibility is
that one has a CFL state as a core of the star, then a shell in
the LOFF state and eventually the exterior part made up of
neutrons. Since in the CFL state the baryonic number is broken
there is superfluidity. Therefore the same mechanism explained
above might work with vortices in the CFL state pinned to the LOFF
crystal. This could reinforce the ideas about the existence of
strange stars (made of up, down and strange quarks). Considering
the very narrow range of values for $\delta\mu$ in order to be in
the LOFF phase one can ask if the previous possibility has some
chance to be realized. Notice that using the typical LOFF value
$\delta\mu\approx .75\Delta_{BCS}$ one would need values of
$\delta\mu$ around $15\div 70~MeV$. Let us consider  a very crude
model of three free quarks with $M_u=M_d=0$, $M_s\not =0$.
Assuming at the core of the star a density around $10^{15}$
g/cm$^3$, that is from 5 to 6 times the saturation nuclear density
$0.15\times 10^{15}$ g/cm$^3$, one finds for $M_s$ ranging between
200 and 300 $MeV$ corresponding values of $\delta\mu$ between 25
and 50 $MeV$. We see that these values are just in the right range
for being within the LOFF window.

\section*{Acknowledgments}
I am grateful to R. Gatto, M. Mannarelli and G. Nardulli for the
very pleasant scientific collaboration on the subjects discussed
in this talk.

\end{document}